\documentclass[journal,comsoc]{IEEEtran}
\usepackage[T1]{fontenc}
\usepackage{color}
\usepackage{bm}
\usepackage{amsthm}

\ifCLASSINFOpdf
\else
\fi
\usepackage{amsmath}
\usepackage[cmintegrals]{newtxmath}
\usepackage{graphicx}
\usepackage{algorithm}
\usepackage{algorithmic}

\usepackage{lineno}
\author{Yunyi Wu and Yongbing Zhang\thanks{Y. Wu, and Y. Zhang are with the Graduate School of Systems and Information Engineering, University of Tsukuba, Japan. E-mail:
wuyunyi0824@outlook.com and ybzhang@sk.tsukuba.ac.jp.}}
\begin{document}

\switchlinenumbers
%
\title{Cost-Efficient Design for 5G-Enabled MEC Servers under Uncertain User Demands}
%
%
%

%

\maketitle

\begin{abstract}
Mobile edge computing (MEC) enhances the performance of 5G networks by enabling low-latency, high-speed services through deploying data units of the base station on edge servers located near mobile users. However, determining the optimal capacity of these servers while dynamically offloading tasks and allocating computing resources to meet uncertain user demands presents significant challenges. This paper focuses on the design and planning of edge servers with the dual objectives of minimizing capacity requirements and reducing service latency for 5G services. To handle the complexity of uncertain user demands, we formulate the problem as a two-stage stochastic model, which can be linearized into a mixed-integer linear programming (MILP) problem. We propose a novel approach called accelerated Benders decomposition (ABD) to solve the problem at a large network scale. Numerical experiments demonstrate that ABD achieves the optimal solution of MILP while significantly reducing computation time.
\end{abstract}



%
\IEEEpeerreviewmaketitle

\section{Introduction}
%
%
%
%
\IEEEPARstart{T}{he} appearance of next-generation network services, such as augmented reality (AR), automated driving, and smart cities, each with significantly different quality of service (QoS) requirements, has introduced many new challenges to current radio access networks (RANs) \cite{Afolabi}. These network services are classified by the International Telecommunication Union (ITU) into three main types: enhanced Mobile Broadband (eMBB), which comprises services with high-bandwidth requirements; massive Machine Type Communications (mMTC), referring to services with relaxed latency and throughput requirements; and ultra-Reliable Low Latency Communications (uRLLC), which focus on the services with high reliability and low latency \cite{TA}. To address these challenges, the Open RAN (O-RAN) Alliance proposed a new RAN architecture consisting of three main components: the radio unit (RU), the distributed unit (DU), and the central unit (CU), as shown in Fig. 1. The link between a DU and an RU is called the fronthaul, while the link between a CU and a DU is referred to as the midhaul \cite{Nakayama}.
 \begin{figure}[!t]
\centering
\includegraphics[width=3.2in]{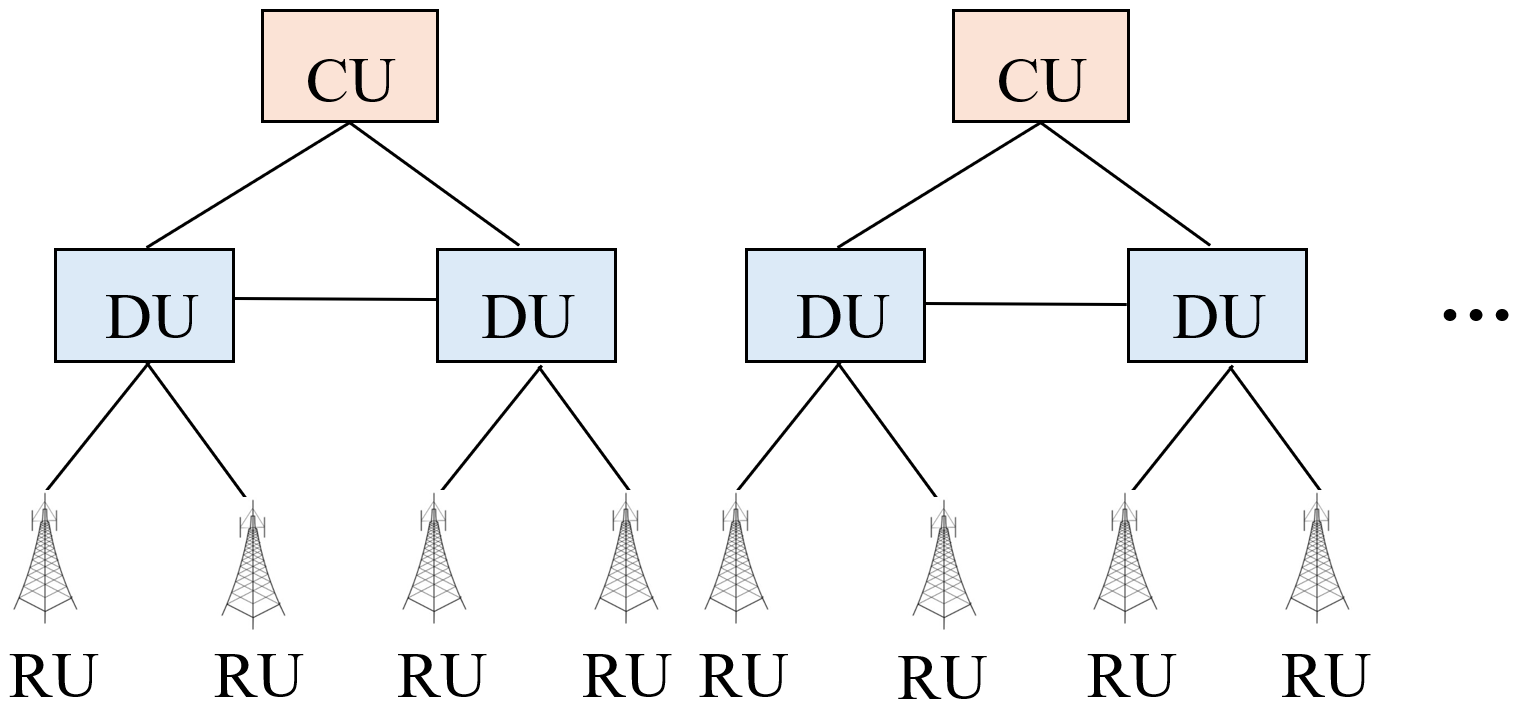}
\caption{An example of O-RAN architecture \cite{Nakayama}.}
\label{fig:2}
\end{figure}

 \begin{figure}[!t]
\centering
\includegraphics[width=1.8in]{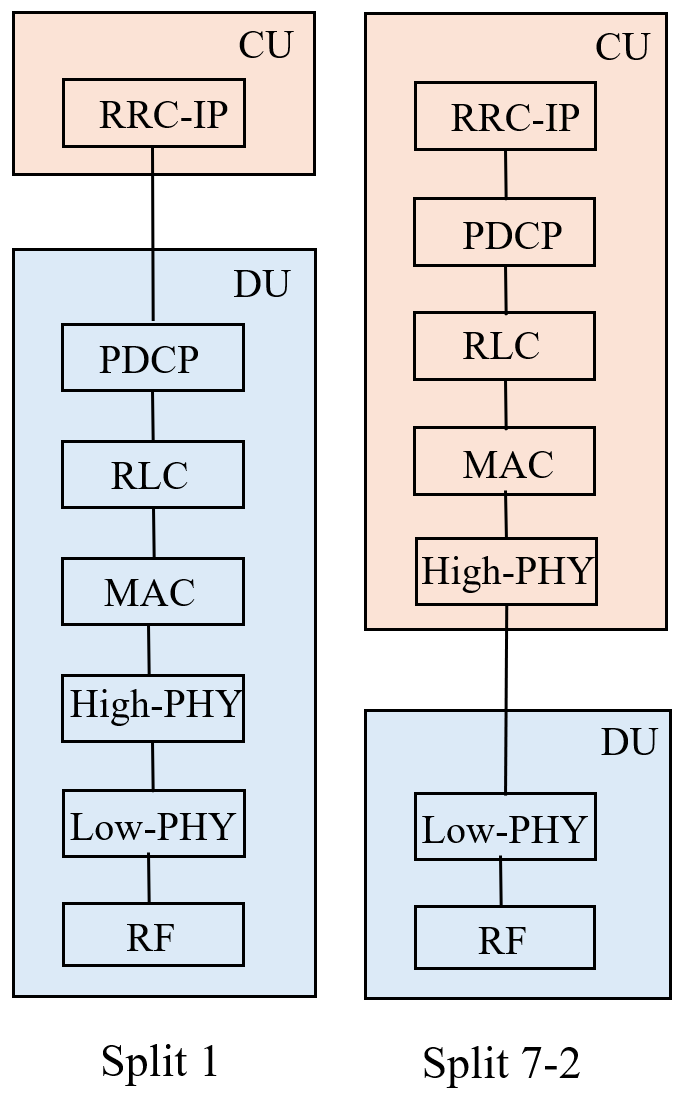}
\caption{An example of two function placement options \cite{3GPP}.}
\label{fig:2}
\end{figure}

In an O-RAN, radio units (RUs) are equipped with antennas responsible for processing the radio frequency (RF) signals from mobile users \cite{Bell}. Distributed units (DUs) are smaller units deployed on mobile edge computing (MEC) servers located close to mobile users, while central units (CUs) are data centers that enable to process vast amounts of data. Both DUs and CUs handle all baseband processing, which includes network functions such as the low and high parts of the physical layer (Low-PHY and High-PHY) processing, Medium Access Control (MAC), Radio Link Control (RLC), Packet Data Convergence Protocol (PDCP) processing, Radio Resource Control (RRC), and Internet Protocol (IP) processing \cite{DOro}. The 3rd Generation Partnership Project (3GPP) proposed eight different options for deploying these functions on DUs or CUs. Fig. 2 illustrates two examples: option split 1 and option split 7-2. In option split 1, most network functions are decentralized and located in DUs, enabling fast responses to guarantee low latency and high reliability for uRLLC demands. Conversely, option split 7-2, where most network functions are shifted toward the CU, is well-suited for handling eMBB traffic with high data rates \cite{3GPP}. O-RAN provides a platform for automated management and dynamic scaling of network functions, enabling quick adjustments to select the most suitable function placement options and meet varying user demands in real time \cite{Marsch}.

Although the computing resources allocated to functions in DUs can be adjusted depending on user demands, it is not possible to scale the DUs during service provision. If the computing resources of DUs are insufficient to host the required functions, user requests should be rejected. Therefore, determining a suitable DU capacity before user traffic arrives is crucial, especially when user traffic patterns are not clear. This paper focuses on the DU design problem of identifying the smallest suitable DU capacity that can accommodate various user demand scenarios. Furthermore, we propose mechanisms for user access and function placement option selection in O-RAN to minimize the latency of network services provided by O-RAN.

The contributions of this paper can be summarized as follows: 1) We formulate the DU design problem, considering uncertain user demands, as a two-stage stochastic problem. 2) We linearize this problem as a mixed-integer linear programming (MILP) model, which can be solved using general optimization solvers. 3) We propose an approach called accelerated Benders decomposition (ABD) to solve the problem at a large network scale. The rest of this paper is organized as follows. Section 2 investigates related work. Sections 3 and 4 introduce the problem considered in this paper and provide its formulation. Section 5 and 6 describe our proposed ABD approach. In Section 7, we compare the performance of our approach with MILP and another approach from previous work. Finally, Section 8 concludes the paper.

\section{Related work}
Many related works focus on designing the DU and CU architecture of O-RAN to enhance RAN flexibility and scalability, following the eight BBU function split options proposed by 3GPP. In these studies, the original BBU functions are modeled as a chain of virtual functions that can be split at designated points, enabling flexible placement of BBU functions between cloud and edge sites. Alabbasi et al. \cite{Alabbasi} consider a three-layer O-RAN architecture, where an RU connects to a DU via a fronthaul, and a DU connects to a CU via a midhaul. To meet user delay requirements, they select split options that place as many functions as possible on DU sites. Liu et al. \cite{Liu} evaluate the computational and link requirements of the midhaul for each split option and propose a network calculus approach to determine the optimal split. Their approach centralizes as many functions as possible on the CU to minimize computational demands on DUs and reduce midhaul bandwidth consumption.

Other studies explore more complex DU designs that offer greater flexibility in BBU function placement. In these models, BBU functions deployed at the edge can be placed not only on a single DU but also on adjacent DUs, provided that user delay requirements are met. Amani et al. \cite{Amani} propose a DU architecture where each DU can connect to multiple RUs. Their approach aims to centralize more BBU functions on fewer DUs while staying within capacity constraints, thereby preserving computational resources for other DUs. Singh et al. \cite{Singh} consider a fixed DU capacity and allow BBU functions from an overloaded DU to be offloaded onto neighboring DUs, preventing capacity bottlenecks.

More studies focus on optimizing BBU function placement on DU sites based on fluctuating user demand. Zhu et al. \cite{Zhu} propose a dynamic approach where, if the traffic load on a DU increases, BBU functions can be migrated to other DUs to prevent overload. Conversely, when traffic decreases, BBU functions can be consolidated onto fewer DUs to minimize the number of active DUs and save energy. Tzanakaki et al. \cite{Tzanakaki} take user delay requirements into account and propose a policy that dynamically adjusts split options as user demand changes, aiming to minimize the deployment cost of functions across cloud and edge sites. However, while previous studies have focused on cost-effective DU utilization by optimizing computational resource usage or minimizing active DUs, they do not consider the cost of deploying DU hardware. In these approaches, DU capacities are fixed and assumed to be large enough to handle all considered request scenarios, without accounting for the impact of varying DU capacity on overall cost efficiency.

Determining the optimal DU capacity is challenging due to the uncertainty of user traffic load and demand. Benders decomposition (BD) is an efficient approach for solving optimization problems with uncertain parameters, such as virtual network function placement and routing problems \cite{Ljubić}\cite{Yu}. In this work, we apply BD to determine the optimal DU capacity while accounting for uncertain user demands and locations. Additionally, we propose an improved BD method that not only accelerates the solution process but also guarantees optimality.
\section{Model description}
 
\begin{table}
\caption{Parameters.}
\label{tab:3}       
\begin{tabular}{ll}
\hline\noalign{\smallskip}
Parameter & Meaning \\
\noalign{\smallskip}\hline\noalign{\smallskip}
$V$ & set of CUs\\
$U$ & set of DUs\\
$R$ & set of RUs\\
$\zeta_{ru}$ & binary indicator: 1 if RU $r$ connects to DU $u$, $r\in R$, $u\in U$\\
$\eta_{uv}$ &  binary indicator: 1 if DU $u$ connects to CU $v$, $u \in U$, $v \in V$ \\
$d_{ru}$ & transmission delay of the fronthaul between RU $r$ and DU $u$,\\&  $r\in R$, $u \in U$\\
$\delta$ & propagation distance-to-delay ratio of a fronthaul\\
$d_1$ & midhaul transmission delay when using function placement\\& option split 7-2\\
$d_2$ & midhaul transmission delay when using function placement\\& option split 1\\
$\sigma$ & computing capacity of a CU\\
$\kappa$ & maximum computing capacity of a DU\\
$f_a$ & computational resource demand of Low-PHY and\\& RF functions\\
$f_b$ & computational resource demand of High-PHY, MAC, RLC,\\& and PDCP functions\\
$f_c$ & computational resource demand of RRC-IP function\\
$I$ & set of users\\
$S$ & set of possible user demand scenarios\\
$\bm{\Phi_{ir}}$ & set of parameters representing the distance\\& between user $i$ and RRH $r$ in each possibility,\\& $\bm{\Phi_{ir}} = \{\phi^{1}_{ir},\phi^{2}_{ir},...,\phi^{|S|}_{ir}\}$, $r\in R$, $i\in I$\\
$\bm{\Pi_{i}}$ & set of parameters representing delay requirement \\& of user $i$ when they request the different service\\& in each possibility, $\bm{\Pi_{i}} = \{\pi^{1}_{i},\pi^{2}_{i},...,\pi^{|S|}_{i}\}$, $i\in I$\\
$\omega_{i}$ & traffic demand of user $i$\\
$\gamma$ & weighting factor for balance two objectives\\ 
\noalign{\smallskip}\hline
\end{tabular}
\end{table}
\begin{table}
\caption{Variables.}
\label{tab:3}       
\begin{tabular}{ll}
\hline\noalign{\smallskip}
Variable & Meaning \\
\noalign{\smallskip}\hline\noalign{\smallskip}
$p_{u}$ & resource capacity of DU $u$, $p_{u}\geq 0$, $r\in R$ \\
$\lambda^{s}_{ir}$ & decision variable that equals 1 if user $i$ accesses RU $r$ in\\& possibility $s$, and equals 0 otherwise, $s\in S$, $r\in R$, $i\in I$ \\
$\theta^{s}_{ru}$ & decision variable that equals 1 if the functions of RU $r$ are\\& placed on DU $u$ in possibility $s$, and equals 0 otherwise,\\&$\theta^{s}_{iu}\in \{ 0,1\}$, $s\in S$, $r\in R$, $u\in U$ \\
$\psi^{s}_{ru}$ & decision variable that equals to 1 if split 7-2 is used for RU $r$ on\\& DU $u$, or equals to 0 if split 1 is used in possibility $s$,\\& $\psi^{s}_{ru}\in \{ 0,1\}$, $s\in S$, $r\in R$, $u\in U$ \\
$d^{s}_{i}$ & latency of the service for user $i$ in possibility $s$,\\& $s\in S$, $r\in R$, $u\in U$, $i\in I$ \\
$x^{s}_{iru}$ & auxiliary variable\\
$y^{s}_{iru}$ & auxiliary variable\\
\noalign{\smallskip}\hline
\end{tabular}
\end{table}
\subsection{O-RAN model}
We consider a 3-layer O-RAN architecture consisting of three main components: CUs, DUs and RUs, denoted by sets $V$, $U$, and $R$. A fronthaul link connects RU $r$ ($r\in R$) and DU $u$ ($u\in U$), denoted by $\zeta_{ru}$, while a midhaul link connects DU $u$ ($u\in U$) and CU $v$ ($v\in V$), denoted by $\eta_{uv}$. The computing capacity of a CU is denoted by $\sigma$ and the maximum computing capacity of a DU is denoted by $\kappa$. This paper examines two function placement options for an O-RAN proposed by 3GPP: split 7-2 and split 1 \cite{3GPP}. As shown in Fig.2, option split 7-2 places Low-PHY and RF functions on the DU, and RRC-IP, PDCP, RLC, MAC, High-PHY functions on the CU, while option split 1 places RRC-IP function on CU, all other functions on the DU. For simplicity, we denote the sum of the computational resource demands of Low-PHY and RF functions as $f_a$, the sum of the computational resource demands of the High-PHY, MAC, RLC and PDCP functions as $f_b$, and the computational resource demand of RRC-IP function as $f_c$. We let $d_1$ represent the midhaul transmission delay provided by option split 7-2, and $d_2$ represent the midhaul transmission delay provided by option split 1. Fronthaul transmission delay determined by the distance $d_{ru}$ between RU $r$ and DU $u$, as well as the propagation distance-to-delay ratio $\delta$.

\subsection{Request model}
Mobile users are represented by the set $I$, and we consider the uncertainty in user locations and user demands. We use the set $\bm{\Phi_{ir}}$ ($i\in I$, $r\in R$) to represent |S| possible user locations, where a specific location possibility is denoted by $\phi^{s}_{ir}$ ($s\in S$). Moreover, user service requests vary over time, with different services having distinct delay requirements. To capture this, we define the set $\bm{\Pi_{i}}$ ($i\in I$) to represent the possible delay requirements for a user $i$ requesting different services. A specific request possibility is denoted by $\bm{\pi^{s}_{i}}$ ($s\in S$), with corresponding traffic demand represented by $\omega_{i}$.

\section{Problem formulation}
5G service providers aim to minimize the cost of deploying the O-RAN while meeting the service demand from mobile users. In this paper, we consider cost savings as minimizing the total DU capacity. Since the capacities of DUs affect the decisions related to function placement for 5G services, which in turn impact the performance of these services (such as latency), determining the optimal DU capacity is challenging, particularly with uncertain user demands. To address this, we consider all possible user locations and demands, aiming to find the optimal DU capacity that meets users' delay requirements across all scenarios. The capacity of DU $u$ ($u\in V$) is denoted by $p_u$. Based on these DU capacity decisions, we also determine which RU a user accesses and the function placement for each possibility to minimize the latency of network services. For a possibility $s$, whether user $i$ accesses RU $r$ is denoted by $\lambda^{s}_{ir}$ ($i\in I$, $r\in R$), and whether the functions of RU $r$ are placed on DU $u$ is denoted by $\theta^{s}_{ru}$ ($i\in I$, $r\in R$), where both $\lambda^{s}_{ir}$ and $\theta^{s}_{ru}$ are binary variables. Additionally, function placement option of RU $r$ on DU $u$ is denoted by $\psi^{s}_{ru}$, and if option split 7-2 is adopted, $\psi^{s}_{ru}$= 1; otherwise, if option split 1 is adopted, $\psi^{s}_{ru}$ = 0. The latency of service for user $i$ ($i\in I$) in possibility $s$ is denoted by $d^{s}_{i}$, and we assume it includes the transmission delays from both the fronthaul and midhaul, as shown below.
\begin{equation}
\begin{split}
d^{s}_{i}=\lambda^{s}_{ir}\theta^{s}_{ru}[\delta d_{ru} + (1-\psi^{s}_{ru})d_1 +  \psi^{s}_{ru} d_2].
\end{split}
\end{equation}
We formulate the problem described above as a two-stage stochastic optimization problem as follows:
\begin{equation}
   \min  \frac{\gamma}{|U|}\sum_{u\in U}p_u + \frac{1}{|S||I|} \sum_{s\in S}\sum_{i\in I}d^{s}_{i},
\end{equation}
subject to
\begin{align}
p_{u} &\leq \kappa, u\in U,\\
\lambda^{s}_{ir} &\leq \phi^{s}_{ir},\ i\in I,r\in R, s\in S,\\
\sum\limits_{r \in R}\lambda^{s}_{ir}&=1,\ i\in I, s\in S,\\
\sum\limits_{u \in U}\theta^{s}_{ru}&= 1,\ r\in R, s\in S,\\
\theta^{s}_{ru} &\leq \zeta_{ru},\ u\in U, r\in R, s\in S,\\
p_u &\geq \sum\limits_{i \in I}\sum\limits_{r \in R}\omega_{i}\lambda^{s}_{ir}\theta^{s}_{ru}  (f_a + (1-\psi^{s}_{ru}) f_b),\nonumber\\&u\in U, s\in S,\\
\sigma &\geq \sum\limits_{i \in I}\sum\limits_{r \in R}\sum\limits_{u \in U}\omega_{i}\lambda^{s}_{ir}\theta^{s}_{ru}\eta_{uv}(\psi^{s}_{ru} f_b + f_c),\nonumber\\ &v\in V, s\in S,\\
d^{s}_{i}&\leq \pi_{i}, i\in I,
\end{align}
\begin{align}
    p_{u}&\in \{0,R+\}, u\in U, \label{const14}\\
    \lambda^{s}_{ir} &\in \{0,1\},\ s\in S, i\in I, r\in R,\label{const14}\\
    \theta^{s}_{ru} &\in \{0,1\},\ s\in S, r\in R, u\in U,\label{const15}\\
    \psi^{s}_{ru} &\in \{0,1\},\ s\in S, r\in R, u\in U,\label{const16}\\
    d^{s}_{i} &\in \{0,R+\},\ s\in S, i\in I,\label{const16}
\end{align}
where $\gamma$ is a weighting factor. Constraint (3) specifies the upper limit of the capacity for each DU. Constraint (4) ensures that a user can only access an RU if it is within the RU's coverage. Constraint (5) enforces that each user can access only one RU. Constraints (6) and (7) ensure that the functions of an RU can only be placed on one DU to which the RU is connected. Constraints (8) and (9) ensure that the functions placed on a DU or CU do not exceed their respective capacities. Finally, Constraint (10) guarantees that the latency of network services satisfies the delay requirements of each user. 

Although constraints (8), (9) and (10) are quadratic constraints, making Problem (2) non-linear, we introduce two auxiliary variables $x^{s}_{iru}$=$\lambda^{s}_{ir}$$\theta^{s}_{ru}$, $y^{s}_{iru}$=$x^{s}_{iru}$$\psi^{s}_{ru}$, along with auxiliary constraints to linearize Problem (2) as a mixed-integer linear programming (MILP) problem. The auxiliary constraints are defined for $i\in I, r\in R, u\in U, s\in S$ as follows:
\begin{align}
-\lambda^{s}_{ir} + x^{s}_{iru} &\leq 0,\\
-\theta^{s}_{ru} + x^{s}_{iru} &\leq 0,\\
\lambda^{s}_{ir} + \theta^{s}_{ru} - x^{s}_{iru} &\leq 1,\\
- x^{s}_{iru} + y^{s}_{iru} &\leq 0,\\
- \psi^{s}_{ru} + y^{s}_{iru} &\leq 0,\\
x^{s}_{iru} + \psi^{s}_{ru} - y^{s}_{iru} &\leq 1,\\
x^{s}_{iru} &\in \{0,1\},\\
y^{s}_{iru} &\in \{0,1\}.
\end{align}

Since Problem (2) is also a virtual function placement problem, which is NP-hard, solving it on a large scale using an optimization solver such as the Gurobi Optimizer \cite{Gurobi} is challenging. Therefore, we propose a more efficient approach to address this problem, which is introduced in the next section.

\section{Benders decomposition for DU design}
Benders decomposition (BD) is an efficient method for solving two-stage stochastic problems and has been widely applied in various fields, such as finance and transportation \cite{Crainic}. BD works by decomposing the original problem into a master problem (MP) and several subproblems (SPs). The process begins by solving the MP, and its solutions are then passed to the SPs. The MP is essentially the original problem with certain constraints removed, meaning its solutions may not be feasible for the full problem. The SPs are used to evaluate the feasibility of the MP’s solutions. If the solutions are not feasible, Benders feasibility cuts are generated and added to the MP to eliminate the infeasible solution region. Conversely, if the solutions are feasible, Benders optimality cuts are generated and added to the MP to update the upper bound (UB) and lower bound (LB) of the original problem. This iterative process continues, alternating between solving the MP and SPs, until the gap between the UB and LB closes, indicating that the optimal solution of the original problem has been found \cite{Geoffrion}. The details of the approach are outlined in Algorithm 1.

In the first iteration $t$ ($t=0$), the upper bound of the solution is set to + $\infty$, and the lower bound is set to - $\infty$. The $MP^{t}$ in the first iteration is defined as a simplified problem that minimizes the DU capacity without considering additional constraints, such as the delay requirements of users. This is formulated as follows:
\begin{equation}
(MP^{t}) \ \ \ \ \  \frac{\gamma}{|U|}\sum_{u\in U}p_u + \frac{1}{|S|} \sum_{s\in S}\alpha_{s}, t=0
\end{equation}
subject to
\begin{align}
p_{u} &\leq \kappa, u\in U, \\
p_{u}&\in \{0,R+\}, u\in U,\\
\alpha_{s} &\in \{0,R+\}, s\in S.
\end{align}
In $MP^{t}$, $\alpha_{s}$ is a variable associated with the optimality cuts in the $SP$s, and in the first iteration $t$ ($t=0$), $\alpha_{s}$ is initialized to 0. We define $|S|$ subproblems, where each SP focuses on minimizing the latency of network services. The formulation of an $SP_s$ ($s\in S$) is given as follows :
\begin{equation}
\label{problem6}
(SP_s) \ \ \ \ \ \ \  \min \frac{1}{|I|}\sum\limits_{i \in I}d^{s}_{i}
\end{equation}
subject to
\begin{equation}
modified \ \ \ \ (4)-(23).
\end{equation}
In modified constraints (4)-(23), the variables $p_u$ ($\forall u\in U$)  are assigned the values obtained from the solutions of the $MP$. In the first iteration, we solve the $MP^{t}$ and use its solutions $\hat{p^{t}_u}$ $(\forall u\in U)$, as input for the $SP$s. For an $SP_s$ ($s\in S$), if it is unfeasible, we add a Benders feasibility cut to $MP^{t}$, which is formulated as follows:

\begin{equation}
\begin{split}
\sum_{i\in I}\sum_{r\in R}\phi^{s}_{ir}w^{0}_{ir}+\sum_{i\in I}w^{1}_{i}+\sum_{r\in R}w^{2}_{r}+\sum_{r\in R}\sum_{u\in U}\zeta_{ru}w^{3}_{r}\\+\sum_{u\in U}p_{u}w^{4}_{u}+\sum_{v\in V}\sigma w^{5}_{v}+\sum_{i\in I}\sum_{r\in R}\sum_{u\in U}\pi^{s}_{i}w^{6}_{inru}\\+\sum_{i\in I}\sum_{r\in R}\sum_{u\in U}w^{9}_{iru}+\sum_{i\in I}\sum_{r\in R}\sum_{u\in U}w^{12}_{iru} \leq 0, s\in S.
\end{split}
\end{equation}
The variables $w^{0}_{ir}$, ..., $w^{12}_{iru}$ are dual variables of a $SP_s$. Furthermore, if $SP_s$ is feasible, we add a Benders optimality cut to $MP^{t}$, which is formulated as follows:
\begin{equation}
\begin{split}
\sum_{i\in I}\sum_{r\in R}\phi^{s}_{ir}w^{0}_{ir}+\sum_{i\in I}w^{1}_{i}+\sum_{r\in R}w^{2}_{r}+\sum_{r\in R}\sum_{u\in U}\zeta_{ru}w^{3}_{r}\\+\sum_{u\in U}p_{u}w^{4}_{u}+\sum_{v\in V}\sigma w^{5}_{v}+\sum_{i\in I}\sum_{r\in R}\sum_{u\in U}\pi^{s}_{i}w^{6}_{inru}\\+\sum_{i\in I}\sum_{r\in R}\sum_{u\in U}w^{9}_{iru}+\sum_{i\in I}\sum_{r\in R}\sum_{u\in U}w^{12}_{iru} \leq \alpha_{s}, s\in S.
\end{split}
\end{equation}

If all $SP$s are feasible, a new upper bound $UB^{'}$ is generated (see steps 13--18). The $MP^{t}$ is updated by incorporating the feasibility and optimality cuts, and the process advances to the next iteration ($t=t+1$). By solving the updated $MP^{t}$, new solutions $\hat{p^{t}_u}$ are obtained and its optimal value is used as the new lower bound (LB) for the problem (see steps 19--24). The iteration continues until the gap between the UB and LB is zero. The solutions $\hat{p_u^{t}}$ from the final iteration represent the optimal DU capacity. Finally, these solutions are used as inputs for the SPs in the last iteration, which are then solved to determine the values of $\lambda^{s}_{ir}$, $\theta^{s}_{ru}$, and $\psi^{s}_{ru}$ (see steps 26--27).

\begin{figure}[!t]
  \label{alg:4}
  \renewcommand{\algorithmicrequire}{\textbf{Input:}}
  \renewcommand{\algorithmicensure}{\textbf{Output:}}
\begin{algorithm}[H]
\caption{Benders decomposition for DU design.}
\begin{algorithmic}[1]
\REQUIRE $V$, $U$, $R$, $I$, $N$, $S$, $UB$, $LB$
        \ENSURE $p_{u}$, $\lambda^{s}_{ir}$, $\theta^{s}_{ru}$, $\psi^{s}_{ru}$
\STATE Set $MP^{t}$ as $MP$ (24), $t=0$
\STATE Set $UB$ = + $\infty$ and $LB$ = - $\infty$
\STATE Solve $MP^{t}$ and achieve solutions $\hat{p_u^{t}}$
\WHILE {$UB$ - $LB$ $> \varepsilon$}
    \FOR{each scenario $s$ ($s\in S$)}
        \STATE Solve the dual problem of $SP_s$ (28)
        \IF{the dual problem of $SP_s$ is unfeasible}
        \STATE Add a feasibility cut (30) to $MP^t$
        \ELSIF{the dual problem of $SP_s$ is feasible}
        \STATE Add an optimality cut (31) to $MP^t$
        \ENDIF
    \ENDFOR
    \IF{all SPs are feasible}
    \STATE $UB^{'}$=$\frac{\gamma}{|U|}\sum_{u\in U}\hat{p^{t}_u} + \frac{1}{|S||I|} \sum_{s\in S}\sum_{i\in I}d^{s}_{i}$,
    \IF{$UB^{'}< UB$ }
    \STATE $UB$ = $UB^{'}$
    \ENDIF
    \ENDIF
    \STATE $t$ = $t$ + 1
    \STATE $MP^{t}$ = $MP^{t-1}$
    \STATE Solve $MP^t$, then obtain a new solution $\hat{p_{u}^{t}}$ and a optimal value $LB^{'}$
    \IF{$LB^{'}> LB$ }
    \STATE $LB$ = $LB^{'}$
    \ENDIF
\ENDWHILE
\STATE Set $p_{u} = \hat{p_u^{t}}$, $\forall u\in U$,
\STATE Input $p_{u}$ into $SP$s into the last iteration, solve $SP$s and obtain $\lambda^{s}_{ir}$, $\theta^{s}_{ru}$, $\psi^{s}_{ru}$
\end{algorithmic}
\end{algorithm}
\end{figure}
\section{Proposed accelerated Benders decomposition}
Since each SP in our DU design problem defined in Section V is an MILP with limited constraints and variables, it can be solved efficiently. Thus, the overall solution speed primarily depends on solving the MP. As numerous Benders cuts are added to the MP in each iteration, we aim to identify and remove redundant cuts to accelerate the solution process. In this section, we introduce a method for detecting these redundant Benders cuts and explain why removing them does not affect the optimality of the solution.

A Benders feasibility cut is added to the MP to eliminate infeasible solutions. If the solution space removed by one feasibility cut is entirely covered by another, the former is redundant and need not be included in the MP. There are such redundant Benders feasibility cuts in $MP^{t}$ (24). For example, consider two Benders feasibility cuts generated from two SPs $s_1$ and $s_2$ ($s_1$, $s_2$ $\in$ $S$). The first cut is given by
\begin{equation}
\sum_{i\in I}\sum_{r\in R}\phi^{s_1}_{ir}w^{0}_{ir}+\sum_{i\in I}\sum_{r\in R}\sum_{u\in U}\pi^{s_1}_{i}w^{6}_{inru}+ \beta \leq 0,
\end{equation}
and the second cut is
\begin{equation}
\sum_{i\in I}\sum_{r\in R}\phi^{s_2}_{ir}w^{0}_{ir}+\sum_{i\in I}\sum_{r\in R}\sum_{u\in U}\pi^{s_2}_{i}w^{6}_{inru}+ \beta \leq 0,
\end{equation}
where 
\begin{equation}
\begin{split}
    \beta = \sum_{i\in I}w^{1}_{i}+\sum_{r\in R}w^{2}_{r}+\sum_{r\in R}\sum_{u\in U}\zeta_{ru}w^{3}_{r}+\sum_{u\in U}p_{u}w^{4}_{u}\\+\sum_{v\in V}c_{v}w^{5}_{v}+\sum_{i\in I}\sum_{r\in R}\sum_{u\in U}w^{9}_{iru}+\sum_{i\in I}\sum_{r\in R}\sum_{u\in U}w^{12}_{iru}.
\end{split}    
\end{equation}
If $\phi^{s_1}_{ir}$ $\leq$ $\phi^{s_2}_{ir}$ and $\pi^{s_1}_{i}$ $\leq$ $\pi^{s_2}_{i}$, then,
\begin{equation}
\begin{split}
\sum_{i\in I}\sum_{r\in R}\phi^{s_1}_{ir}w^{0}_{ir}+\sum_{i\in I}\sum_{r\in R}\sum_{u\in U}\pi^{s_1}_{i}w^{6}_{inru}+ \beta \leq\\ \sum_{i\in I}\sum_{r\in R}\phi^{s_2}_{ir}w^{0}_{ir}+\sum_{i\in I}\sum_{r\in R}\sum_{u\in U}\pi^{s_2}_{i}w^{6}_{inru}+ \beta \leq 0,
\end{split}
\end{equation}
thus, if Benders feasibility cut (33) holds, cut (32) will also be valid. In this case, cut (32) is entirely covered by cut (33), making it redundant and removable from $MP^{t}$ (24). We classify such Benders feasibility cuts, like cut (32), as redundant cuts. To improve computational efficiency, we identify and remove these redundant cuts in step 8 of Algorithm 1.

\section{Numerical experiments and performance evaluation}

\begin{table}
\caption{Parameters settings.}
\label{tab:3}       
\begin{tabular}{ll}
\hline\noalign{\smallskip}
Parameters & Value \\
\noalign{\smallskip}\hline\noalign{\smallskip}
traffic rate of eMBB & 20 Mb/s\\
traffic rate of mMTC & 1 Mb/s\\
traffic rate of uRLLC & 5 Mb/s\\
delay requirement of eMBB & 100 ms\\
delay requirement of mMTC & 100 ms\\
delay requirement of uRLLC & 10 ms\\
computational resource demand of $f_a$ & 2 RCs per Mb/s\\
computational resource demand of $f_b$ & 4 RCs per Mb/s\\
computational resource demand of $f_c$ & 1 RCs per Mb/s\\
maximum computing capacity of a DU $\kappa$ & 4096 RCs\\
computing capacity of a DU $\sigma$ & 32768 RCs\\
transmission delay for split 7-2 $d_1$ & 0.25 ms\\
transmission delay for split 1 $d_2$ & 30 ms\\
distance between a DU and a RU $d_{ru}$ & 0.5-4.0 km\\
propagation delay-to-distance ratio $\delta$ & 10 $\mu$s/km\\
\noalign{\smallskip}\hline
\end{tabular}
\end{table}
We consider three network models in the numerical experiment (1) 2 CUs, 4 DU, and  8 RUs; (2) 4 CUs, 8 DUs, and 16 RUs; (3) 8 CUs, 16 DUs and 32 RUs. The network topology follows a similar structure to Fig. 1, where each CU connects to 2 DUs, and each DU connects to 2 RUs. Additionally, two DUs connected to the same CU are interconnected, allowing RU functions to be placed not only on their directly connected DU but also on a neighboring DU. Following \cite{Ojiaghi}, the computational resource demands for functions are defined as $f_a$ = 2, $f_b$ = 4, $f_c$ = 1 CPU reference core per Mb/s. The computing capacity of a CU is set to 32,768 CPU cores, while the maximum computing capacity of a DU is 4,096 CPU cores. According to \cite{Chitimalla}, the midhaul transmission delays for function split options 7-2 and 1 are set to 0.25 ms and 30 ms, respectively. The distance between a DU and an RU is randomly generated within the range of 0.5 km to 4 km, with a propagation delay-to-distance ratio of 10 µs/km, as defined in \cite{Garcia}. User locations are randomly distributed, allowing a user to be within the coverage of multiple RUs. Users can request three types of network services: eMBB, mMTC, and uRLLC. The traffic rates and delay requirements for each service are summarized in Table III. 

We compare our proposed accelerated Benders decomposition approach (ABD) with three baseline approaches: (1) directly solving the problem using Gurobi, labeled as MILP; (2) applying conventional Benders decomposition \cite{Geoffrion}, labeled as BD; and (3) minimizing network service delay under a fixed DU capacity, where each DU's capacity is predefined to ensure no user requests are rejected \cite{Zarandi}, labeled as "Fix-DU". The comparison experiments were conducted on a Windows 10 computer with 64 GB of memory.

\subsection{Feasibility analysis}
\begin{table}
\caption{Total cost comparison of four approaches across different network models.}
\centering
\begin{tabular}{|c|c|c|c|c|}\hline
 & MILP & BD & ABD & Fix-DU\\ \hline
2-CU network, 100 users & 6.17 & 6.17 & 6.17 & 41.22\\ \hline
4-CU network, 500 users & 16.13 & 16.13 & 16.13 & 41.23\\ \hline
8-CU network, 1000 users& - & 34.79 & 34.79 & 41.25\\ \hline
\end{tabular}
\end{table}

We compare four approaches across three network models: a 2-CU network with 100 users, a 4-CU network with 500 users, and an 8-CU network with 1,000 users. A total of 500 possible user demand scenarios are considered, with the weight $\gamma$ set to 0.01 to balance DU capacity and network service delay. Table IV shows that for smaller network scales, both BD and ABD achieve the optimal solutions, identical to those obtained by MILP. However, as the network scale increases to the 8-CU network, Gurobi fails to solve the problem due to a "Out of Memory", as it cannot input all variables and constraints. In contrast, BD and ABD continue to obtain the optimal solution by decomposing the large problem into an MP and SPs, which are significantly smaller and easier to solve. Additionally, BD and ABD consistently achieve lower total costs compared to the "Fix-DU" approach. The total costs for "Fix-DU" remain nearly constant across all network models because DU capacities in this approach are fixed and sufficiently large to provide low-latency service for almost all users at the edge.

\subsection{Performance evaluation}
\begin{figure}[!t]
\centering
\includegraphics[width=3.5in]{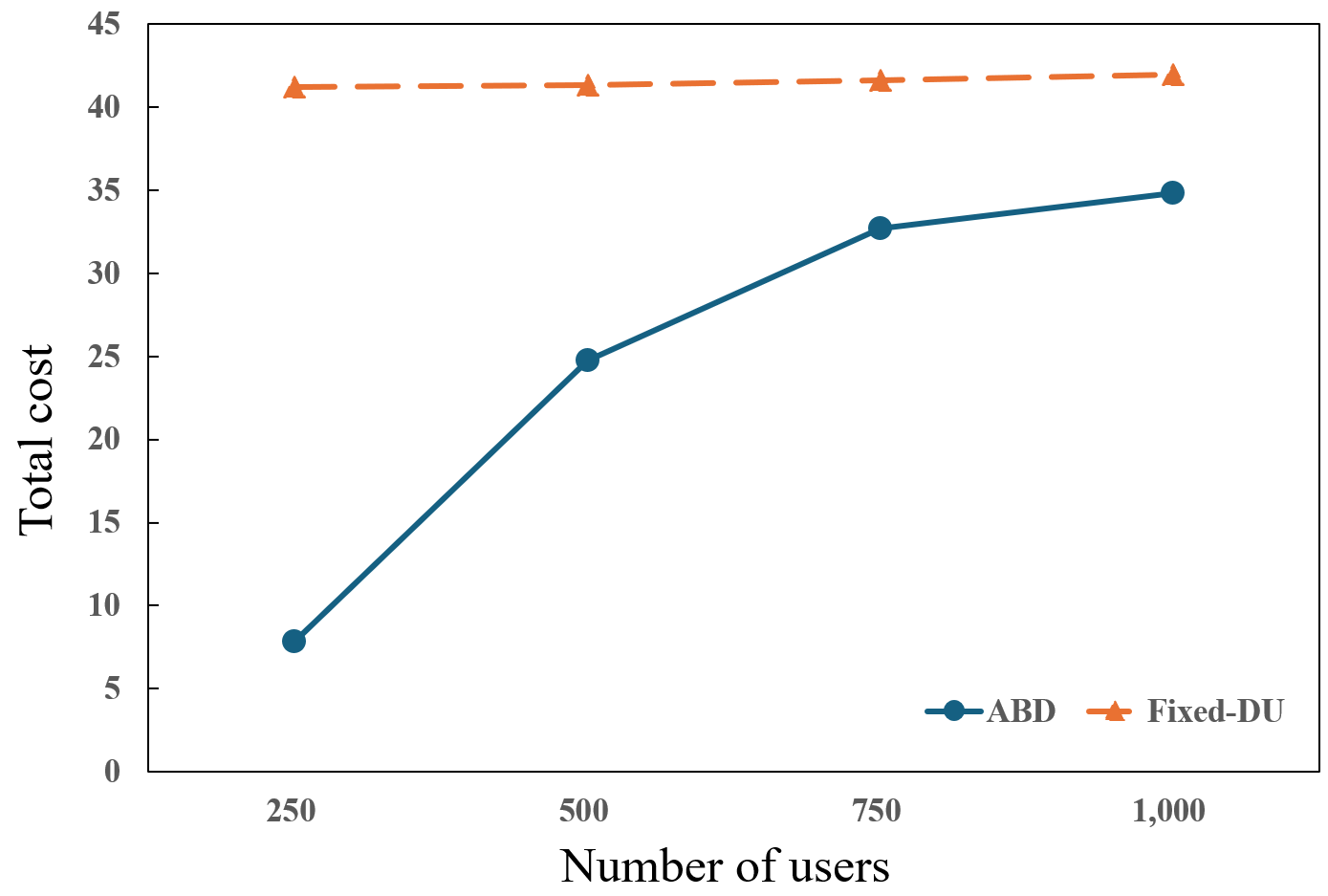}
\caption{Total cost comparison between ABD and Fix-DU in an 8-CU network with an increasing number of users.}
\label{fig:4}
\end{figure}

We next compare the total costs obtained by the BD, ABD, and "Fix-DU" approaches in an 8-CU network as the number of users increases from 250 to 1,000. The number of possible user demand scenarios is set to 1,000. Since the results of BD and ABD are identical, only the results for ABD and "Fix-DU" are shown in Fig. 3. As the number of users increases, the total cost of ABD also increases due to the greater DU capacities required. To accommodate the growing traffic demands, more functions need to be placed on DUs, which increases their capacity. However, the rate of increase in the total cost of ABD slows down as the user count reaches 1,000. This is because ABD transfers more functions from the DUs to the CUs when user delay requirements are met, preventing a significant increase in DU capacities. Compared to simply increasing DU capacity, this strategy leads to a lower overall cost.

\begin{figure}[!t]
\centering
\includegraphics[width=3.5in]{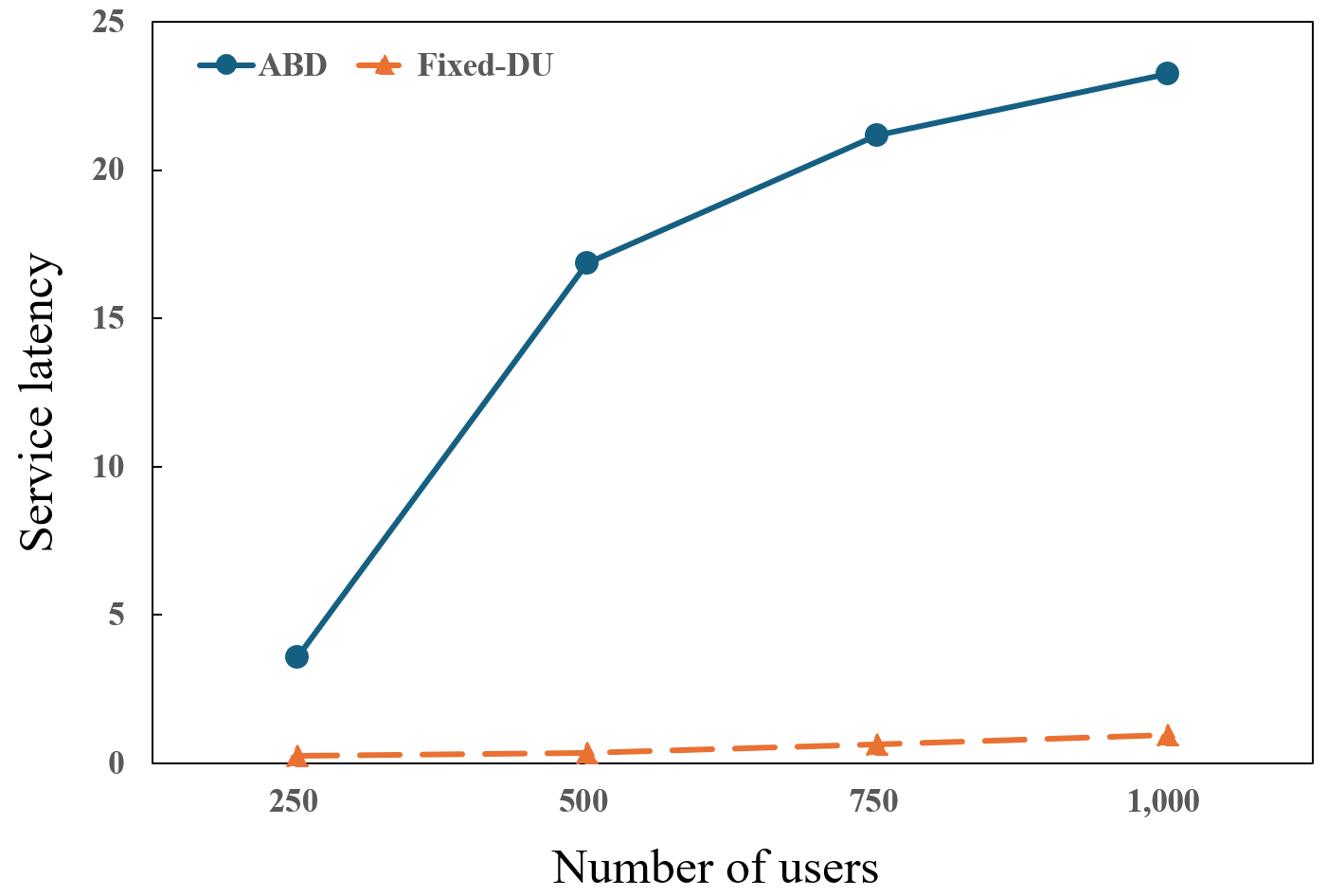}
\caption{Comparison of average service latency between ABD and Fix-DU in an 8-CU network with an increasing number of users.}
\label{fig:4}
\end{figure}

Fig. 3 also shows that the total cost of "Fix-DU" increases only slightly as the number of users grows. This is because the approach pre-allocates a sufficiently large DU capacity to handle traffic, regardless of whether the traffic rate is high or low. The advantage of this strategy is that as many functions as possible are placed at the edge, resulting in very low service latency, as shown in Fig. 4. Compared to ABD, the average service latency of "Fix-DU" remains consistently below 1 ms. However, this approach leads to inefficient computational resource usage when traffic rate is low. Even under high traffic conditions, services such as eMBB and mMTC are not highly delay-sensitive and can be processed at the CU, meaning excessive DU capacity is unnecessary. Determining an appropriate DU capacity is particularly challenging given the uncertainty in user traffic demand, and our proposed approach effectively addresses this issue.

\begin{figure}[!t]
\centering
\includegraphics[width=3.5in]{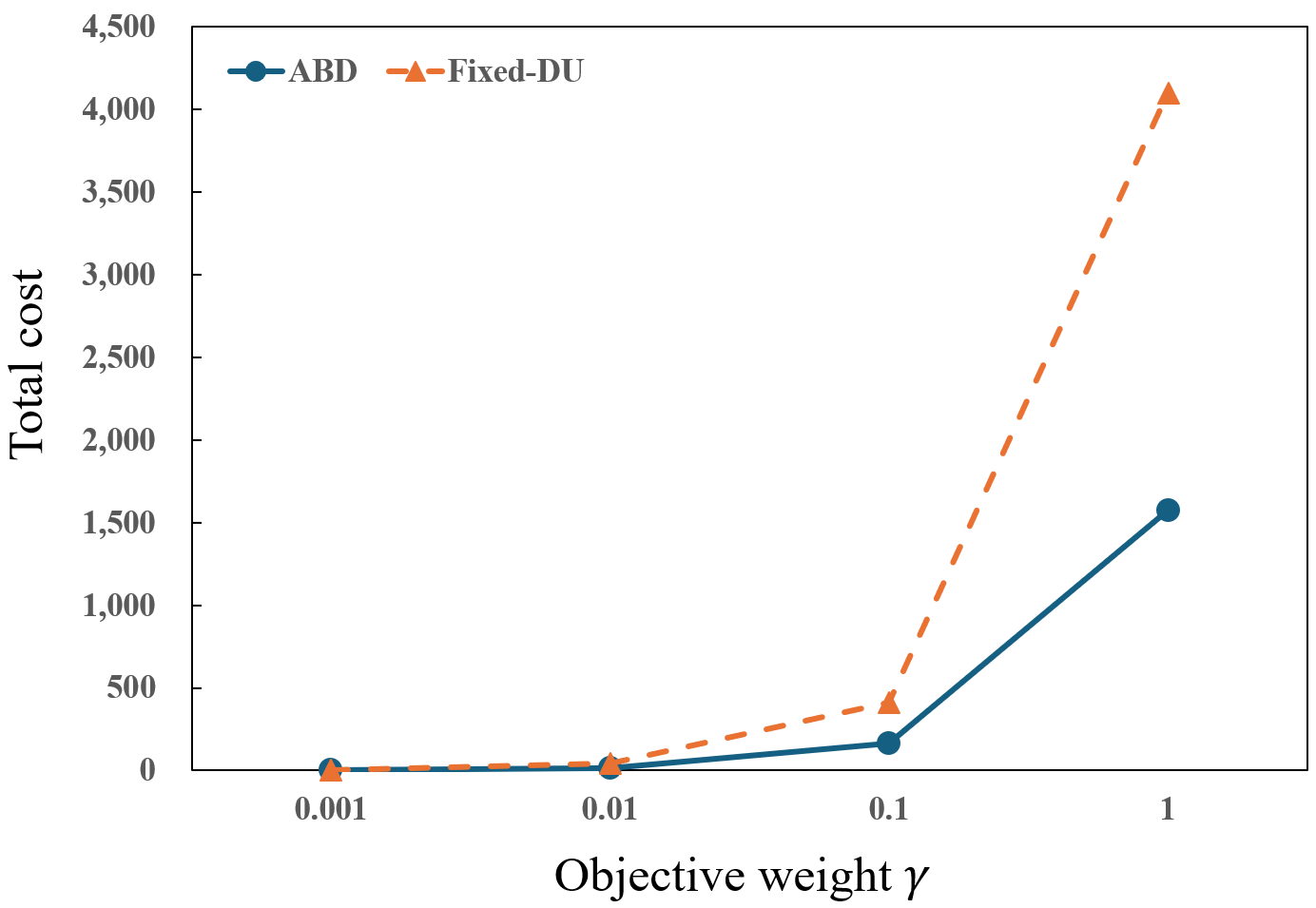}
\caption{Comparison of total cost between ABD and Fix-DU with varying objective weight $\gamma$.}
\label{fig:4}
\end{figure}

Moreover, our approach can adjust the policy based on the network service’s focus, whether on minimizing DU capacity or reducing service latency. Fig. 5 illustrates the total cost for different values of the objective weight $\gamma$. A lower value of $\gamma$ represents service provider focusing on minimizing service latency more, while the lower weight $\gamma$ indicates that the service provider prioritizes minimizing service latency, while a higher value of $\gamma$  reflects a focus on minimizing DU capacity. As shown in Fig. 5, when the focus is on minimizing service latency, our approach achieves a total cost similar to that of "Fix-DU". However, when the emphasis shifts toward minimizing DU capacity, our approach results in a significantly lower total cost compared to "Fix-DU".

\subsection{Comparison of computing time}
\begin{figure}[!t]
\centering
\includegraphics[width=3.5in]{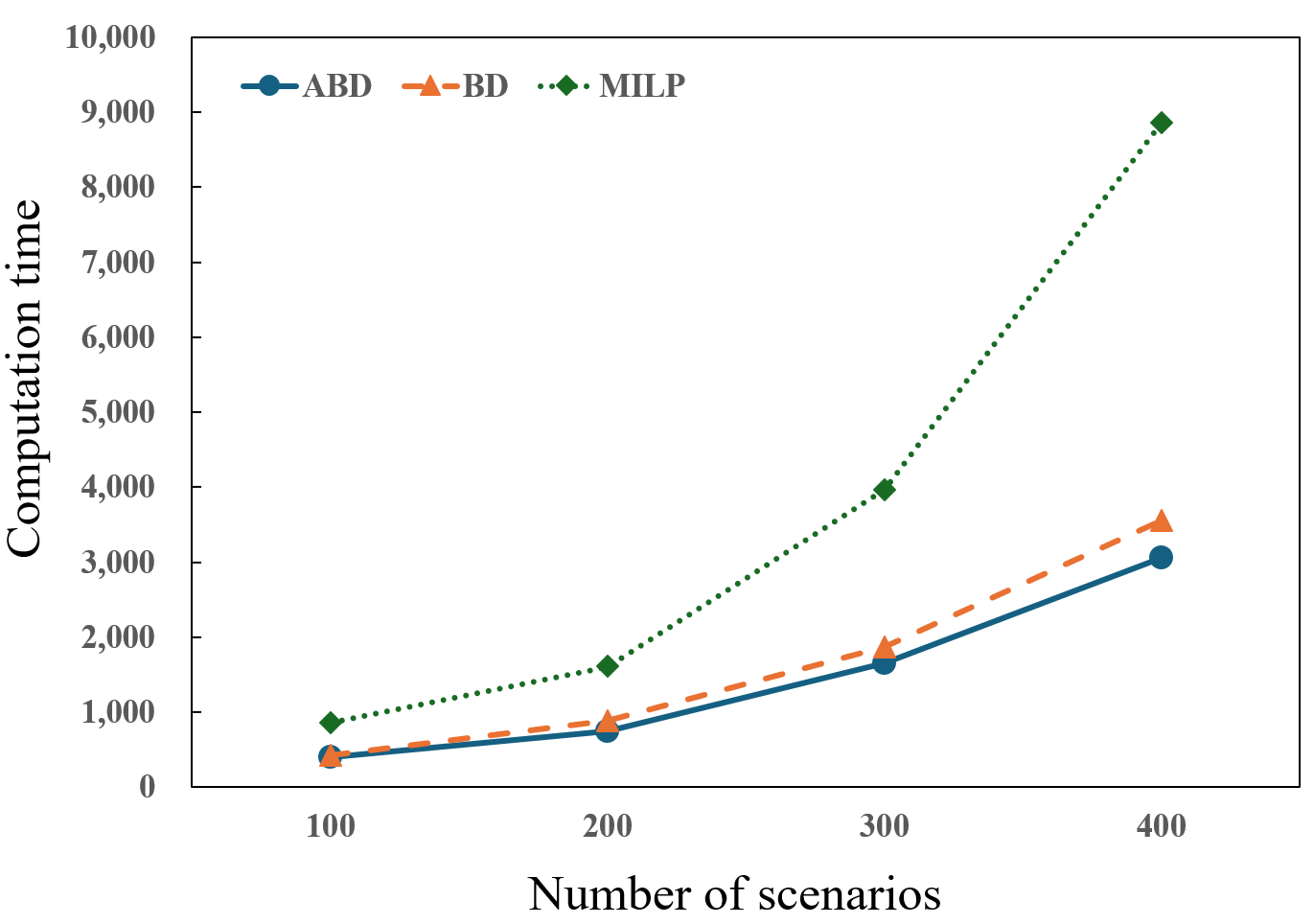}
\caption{Comparison of computation time for MILP, BD, and ABD as the number of user scenario increases (4-CU network).}
\label{fig:4}
\end{figure}
We compare the computation time of three approaches: MILP, BD, and ABD, all of which can obtain the optimal solution to the problem in an 4-CU network. The number of users is set to 500. As shown in Fig. 6, the computation time of the proposed ABD approach is consistently the lowest among the three, as number of possible user demand scenarios increases from 100 to 400. In contrast, the computation time for MILP is significantly higher than both BD and ABD. Additionally, the rate of increase in computation time is much steeper for MILP as the number of users grows.

\begin{figure}[!t]
\centering
\includegraphics[width=3.6in]{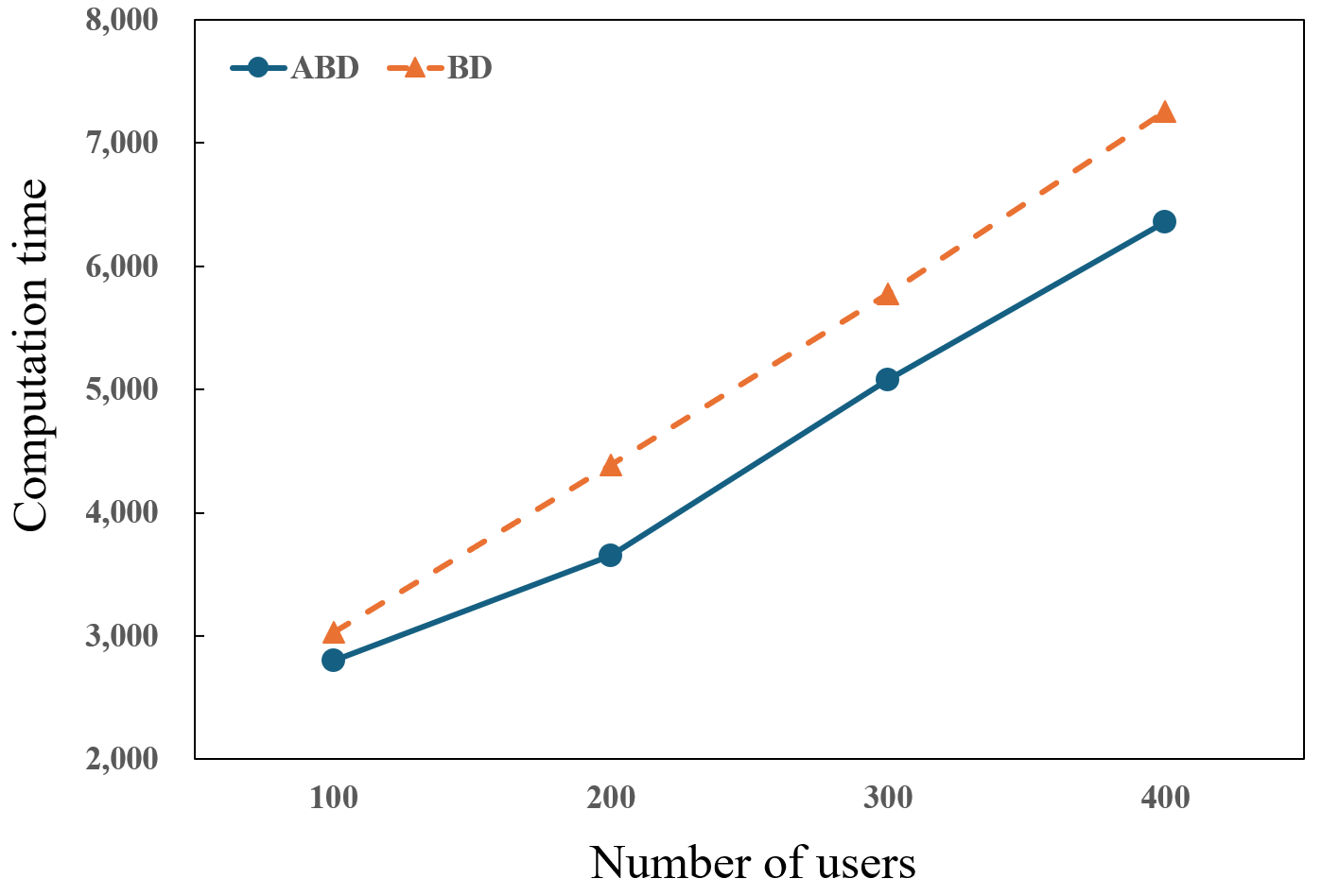}
\caption{Comparison of computation time between BD and ABD as the number of users increases (8-CU network).}
\label{fig:4}
\end{figure}

\begin{figure}[!t]
\centering
\includegraphics[width=3.6in]{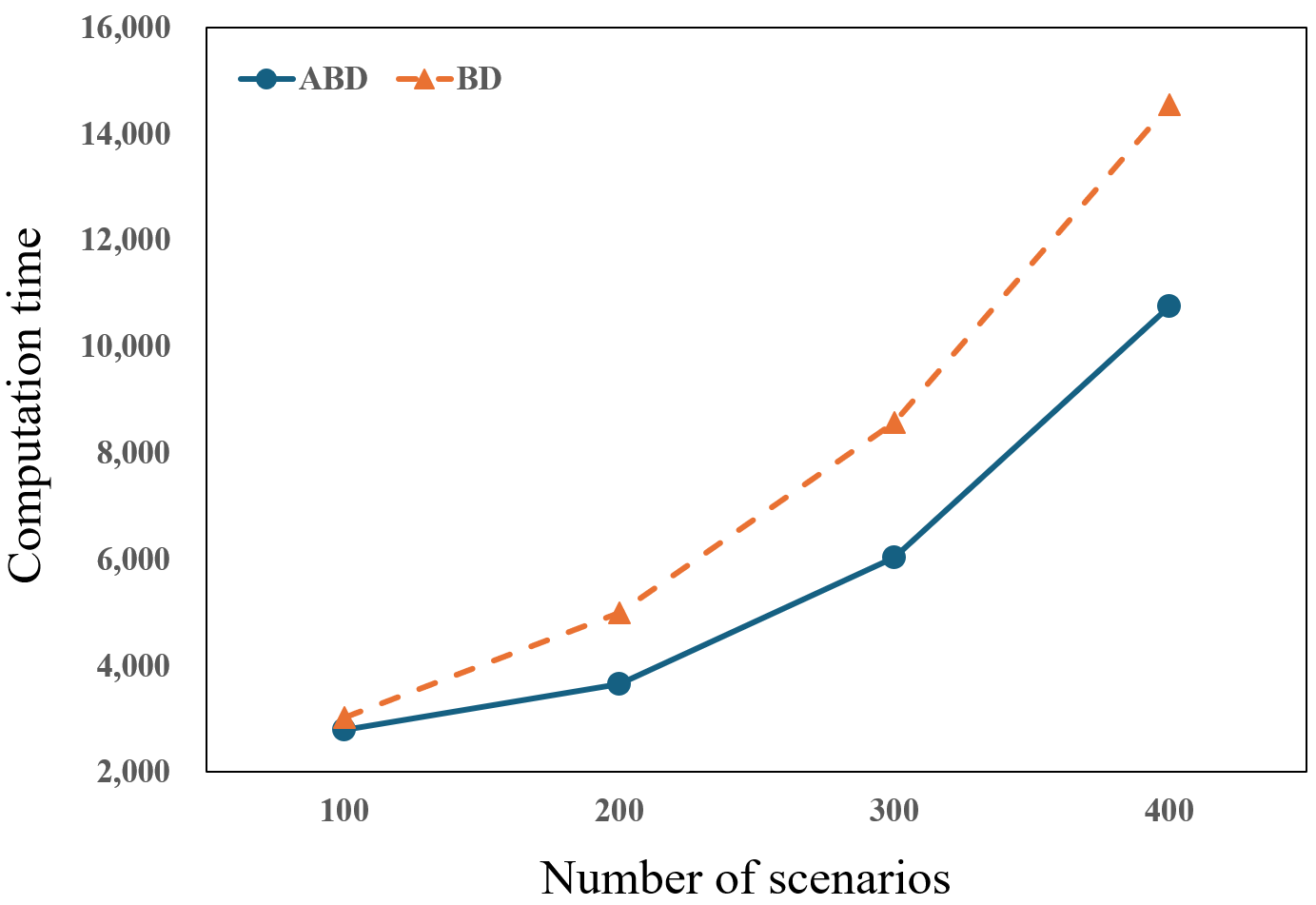}
\caption{Comparison of computation time between BD and ABD as the
number of user scenario increases (8-CU network).}
\label{fig:4}
\end{figure}

Since MILP cannot solve the problem in an 8-CU network, we compare the computation time of BD and ABD in this network. Fig. 7 shows the computation time of the two approaches as the number of users increases from 100 to 400, while Fig. 8 shows the computation time as the number of possibilities increases from 100 to 4000. The results indicate that the computation time of our proposed ABD is consistently faster than BD. In particular, in scenarios with higher numbers of possibilities, ABD is significantly faster than BD, with the speed gap between the two approaches being 26\%.

\section{Conclusion}
In this paper, we focus on the design of an MEC edge server for O-RAN, considering uncertain user locations and demands. The goal of the design is to minimize both the capacity of edge servers and service latency. We first formulate the problem as a two-stage stochastic problem, which is also an MILP problem. Since solving the MILP problem is challenging at large network scales, we propose an approach inspired by the conventional BD method. This approach decomposes the original problem into an MP, which determines server capacity, and several SPs that decide user access and network function placement. The MP and SPs are solved individually, and Benders cuts are generated from the SPs and added to the MP to reduce the solution space until the optimal solution is found. Additionally, we propose an approach called ABD, which removes redundant Benders cuts from the MP to accelerate the overall solving process. Numerical experiments show that our proposed ABD approach can obtain the optimal solution, similar to BD and MILP, but with much lower computation time.


%
\section*{Acknowledgments}
This work has been partially supported by JSPS Grant-in-Aid for Scientific Research (C) 21K04544.

\ifCLASSOPTIONcaptionsoff
\newpage
\fi

\end{document}